\documentclass[10pt,reprint]{socc15}

\usepackage{amsmath}
\usepackage{mathptmx}
\usepackage{latexsym}

\usepackage{epstopdf}
\usepackage{footnote}
\usepackage{listings}

\usepackage{xcolor}
\usepackage{multirow}
\usepackage{graphicx}
\usepackage{subfigure}
\usepackage{url}
\usepackage{xfrac}
\usepackage{balance}
\usepackage{srcltx}
\usepackage{xspace}
\usepackage{booktabs}
\usepackage{enumitem}

\usepackage{soul} 

\usepackage{algorithm}
\usepackage{algpseudocode}
\usepackage{float}
\newfloat{algorithm}{t}{lop}
\graphicspath{{figs/}}

\newcommand{\rr}[0]{\textsf{RR}\xspace}
\newcommand{\sr}[0]{\textsf{SR}\xspace}
\newcommand{\rw}[0]{\textsf{RW}\xspace}
\newcommand{\sw}[0]{\textsf{SW}\xspace}

\usepackage[breaklinks]{hyperref}
\hypersetup{pdfborder = 0 0 0}
\hypersetup{pdftitle={}}
\hypersetup{pdfauthor={}}

\begin{document}

\setlength{\pdfpageheight}{\paperheight}
\setlength{\pdfpagewidth}{\paperwidth}

\conferenceinfo{Extended version of the paper to appear in SoCC'15}{} 
\copyrightyear{2015} 
\copyrightdata{978-1-nnnn-nnnn-n/yy/mm} 
\doi{nnnnnnn.nnnnnnn}

\title{Forecasting the cost of processing multi-join\\
queries via hashing for main-memory databases\\(Extended version)}

\authorinfo{Feilong Liu}
		{Computer Science \& Engineering\\The Ohio State University}
		{liu.3222@osu.edu}
\authorinfo{Spyros Blanas}
		{Computer Science \& Engineering\\The Ohio State University}
		{blanas.2@osu.edu}

\maketitle

\begin{abstract}

Database management systems (DBMSs) carefully optimize complex
multi-join queries to avoid expensive disk I/O. 
As servers today feature tens or hundreds of gigabytes of RAM, a
significant fraction of many analytic databases becomes memory-resident.
Even after careful tuning for an in-memory environment, 
a linear disk I/O model such as the one implemented in
PostgreSQL may make query response time predictions that are up to
$2\times$ slower than the optimal multi-join query plan over
memory-resident data.
This paper introduces a memory I/O cost model to identify good evaluation
strategies for complex query plans with multiple hash-based equi-joins
over memory-resident data.
The proposed cost model is carefully validated for accuracy using three
different systems, including an Amazon EC2 instance,
to control for hardware-specific
differences. 
Prior work in parallel query evaluation has advocated right-deep and bushy
trees for multi-join queries due to their greater parallelization and
pipelining potential. 
A surprising finding is that the conventional wisdom
from shared-nothing disk-based systems does not
directly apply to the modern shared-everything memory hierarchy. 
As corroborated by our model, the performance gap between the optimal
left-deep and right-deep query plan can grow to about $10\times$ as the
number of joins in the query increases.

\end{abstract}

\section{Introduction}

Ad-hoc analytical queries commonly perform a series of equi-joins
between different tables.
A database management system needs to accurately forecast the
response time of different join evaluation orders
to provide timely answers to these queries.
A major component of this prediction is an estimation
of the cost to access the data. These I/O cost models have traditionally
assumed that all data reside in a single directly-attached hard disk.
However, cloud infrastructure providers such as Amazon EC2 offer memory-optimized
instance types with dozens of vCPUs and hundreds of gigabytes of memory.
In this environment, a significant fraction of the database fits in
memory and must be processed in parallel by multiple CPU cores for
timely responses to user queries.

The database research community has been investigating 
query processing techniques for memory-resident datasets for more than 
three decades
\cite{dewitt84, lehman_carey_sigmod1986} and has 
developed detailed cost models for single-core CPUs with deep cache
hierarchies
\cite{manegold2002generic,manegold00}.
The commoditization of multi-core, multi-socket systems have rekindled
research interest in query execution \cite{cjoin09, krikellas_viglas}
and efficient processing techniques for modern hardware
\cite{yinan_ippokratis, polychroniou14}.
In-memory hash-based join algorithms, in particular, have received
significant attention \cite{cagri_sortorhash, blanas:memory,
manycore_join_sigmod14}.
The ramifications of ample single-node parallelism for multi-join
query evaluation over in-memory data, however, are not as well understood. 

As a consequence, systems implementers find themselves in the dark
regarding how to best synthesize query plans for complex multi-join
queries that efficiently use all CPU cores, and turn to prior research in
parallel query evaluation for parallel databases with a shared-nothing
architecture. 
A multi-join query is commonly evaluated as a series of hash-joins between two
tables. 
For every hash join, the query optimizer chooses which input will be
stored in the hash table (henceforth referred to as the \emph{build} or
\emph{left} side) and which input will be used to probe the hash table
(henceforth referred to as the \emph{probe} or \emph{right} side).
Prior work in parallel hash-based multi-join query evaluation has
advocated right-deep query trees, because all build (``left'') subtrees can be
processed in parallel and then probed in a single long pipeline
\cite{dono:tradeoff}.
This conventional wisdom on good query plans is commonly encoded in
ad-hoc heuristics in parallel query optimizers to avoid the combinatorial
explosion of the search space \cite{Ono90}.
Although a single
database server may have as many CPU cores as a parallel database in the
past, the hardware architecture does not otherwise resemble a
shared-nothing environment.
Existing query optimizers need to be augmented to accurately model the
resource needs of ad-hoc queries 
over in-memory data.

This paper focuses on hash-based join evaluation strategies for main-memory
databases and contributes a detailed cost model that accurately
predicts the query response time of ad-hoc multi-join queries.
Our thesis is that the response time for processing a complex multi-join
query using all CPU cores is proportional to the number of memory
operations weighted by the performance cost for different access
patterns.
By exhaustively exploring all possible query plans for a join between 4
tables, we find that the memory access cost that is predicted by our
model proves to be an accurate proxy for elapsed time. 
We then demonstrate that linear disk I/O models, such as
the current cost model in PostgreSQL, cannot accurately account
for the access patterns to and from main memory even after optimally
adjusting the model weights through linear regression.

The proposed cost model is conceptually simple as it does not 
account for the multi-layered cache hierarchy or any NUMA
effects. We find that the loss of accuracy due to these two factors is
minor for the cache-oblivious non-partitioned
hash join algorithm that we model.
Yet by only focusing on the top layer of the memory hierarchy, 
our model gains in versatility and can readily adapt to the
underlying hardware. 
The conceptual simplicity of the model is particularly important in
light of the limited topological information about the CPU and cache
hierarchy that can be gleaned from a VM in a cloud environment.
Experiments with on-premises systems and
Amazon EC2 show that our model accurately predicts how the response time
of individual query plans will change based on the performance
characteristics of different systems.  

A surprising finding from our thorough experimental evaluation is that
some left-deep query trees can be more efficient than their bushy and
right-deep tree counterparts if the join pipeline must run until
completion because they result in substantially less memory traffic
during execution.
As corroborated by our model, the performance gap between the optimal
left-deep and right-deep query tree can grow to about $10\times$ as the
number of joins in the query increases.
This finding challenges the widely held belief among systems
implementers that MPP-style optimization is ``good enough'' for
parallel in-memory query processing.

\section{Background and Related Work}
\label{sec:relwork}

\subsubsection*{Query processing techniques for in-memory data}
Research in main memory database management systems commenced more than
three decades ago. Early work investigated join algorithms for main
memory database systems by
DeWitt et al.~\cite{dewitt84} and an investigation of in-memory query
processing techniques by Lehman and Carey
\cite{lehman_carey_sigmod1986}. 
As the CPU architecture evolved, additional hardware operations were
shown to become performance bottlenecks.
Ailamaki et al.~\cite{ailamaki99} studied
the performance bottlenecks of four commercial
database management systems and highlighted the importance of minimizing
L2 data cache stalls and L1 instruction cache stalls.
Shatdal et al.~\cite{shatdal94} proposed to make the join algorithm
cache-conscious to improve locality and join performance by adding a
partitioning step before the join. 
Boncz et al.~\cite{boncz99} proposed radix-based query processing
techniques to improve in-memory performance by reducing cache and TLB
misses.

As multi-core CPU architectures became the norm, the database research
community explored alternatives to the established query
processing paradigm.
Harizopoulos et al.~\mbox{\cite{harizopoulos05_qpipe}} proposed
a pipelined model for query processing where operators map to hardware
cores and can be shared among multiple queries.
Arumugam et al.~\mbox{\cite{datapath}} developed a push-based dataflow
engine for query
processing. Krikellas et al.~\mbox{\cite{krikellas_viglas}} investigated
techniques to dynamically generate code for query execution. 
Neumann \mbox{\cite{neumann_llvm}} proposed to further improve
performance by compiling queries into native machine
code using LLVM.
Giannikis et al.~\mbox{\cite{SharedDB}} designed a system to process
thousands of concurrent queries by sharing computation and intermediate
results.

There is also a lively and ongoing debate on efficient parallel in-memory equi-join
techniques.
Balkesen et al.~\mbox{\cite{cagri_sortorhash}} compare sort-merge and
radix-hash joins in a multi-core, main-memory environment and claimed that
radix-hash join was superior to sort-merge join.
Leis et al.~\mbox{\cite{manycore_join_sigmod14}} break input data into
small fragments and assign each to a thread to run entire
operator pipelines. By controlling the dispatch of data
fragments, query execution can be parallelized
dynamically in a NUMA-aware fashion.
Polychroniou and Ross \mbox{\cite{polychroniou14}} proposed a family of
main-memory partitioning algorithms that includes hash, radix and range
partitioning.
Li et al.~\mbox{\cite{yinan_ippokratis}} studied data shuffling in a
NUMA system and improved the performance by employing
thread binding, NUMA-aware thread allocation and relaxed global
coordination.
Lang et al.~\mbox{\cite{langimdm13}} optimized the hash join for NUMA
architectures by optimizing the physical representation of the hash
table.
Barber et al.~\cite{Barbervldb15} introduced concise hash
tables which significantly reduce memory consumption while
still offering competitive performance.

Researchers also have proposed cost models for in-memory data
processing.
Manegold et al.~\cite{manegold2002generic}
proposed a hierarchical model to capture all levels of a
cache hierarchy and show that this can accurately capture the total
response time of quick sort, hash join and
partitioning operations after calibrating for the CPU cost of each
algorithm.

\subsubsection*{Parallel query evaluation strategies}

Multi-join query plan optimization is a difficult problem because of the
uncertainty over the cardinality of intermediate results and
the large optimization space. 
Chaudhuri summarized the foundations and significant research
findings of query optimization \cite{Chaudhuri98}, and Lohman has
recently identified open problems in query
optimization
\cite{is_qo_a_solved_problem}.

Prior research has extensively studied parallel and distributed
query optimization techniques.
R* was an early research prototype that optimized queries to minimize
data transfer costs in a shared-nothing environment \cite{lohman85}.
Chen et al.~\cite{chen:sched} proposed to generate parallel query plans
in two phases: first generate the optimal query plan for a single 
node, and then to parallelize it for a multi-processor environment.  
Hong and Stonebraker \cite{hong:optimization} showed that two phase
optimization produces good query plans for shared-memory systems.
Researchers have also studied parallel hash join algorithms.
DeWitt and Gerber \cite{dewitt:multiprocessor} introduced multi-processor
versions of popular hash join algorithms for a single join and evaluated
their advantages and disadvantages in a multi-processor environment.
Recently,
Chu et al.~\mbox{\cite{MagdaLFjoin15}} propose a novel distributed
multi-way join algorithm that has been theoretically proven to have better
performance than evaluating the multi-way join as a series of binary joins.

The performance of different multi-join evaluation strategies has also
been extensively studied.
Schneider and DeWitt \cite{dono:tradeoff} studied multi-way join query
processing through hash-based methods in the Gamma parallel database
system. 
They observed that trees with shallow hash-join build phases and a long
probe pipeline (``right-deep'' trees) can provide significant
performance advantages in large multiprocessor database machines. 
The key advantages of right-deep query trees are the ample
parallelization potential, as each build subtree can be executed
concurrently, and that no intermediate results need to be materialized
in the long probe pipeline because results are streamed from one
operator to the next.

Right-deep trees introduce the problem of processor allocation to each
join sub-tree during parallel evaluation, and 
Lu et al.~\cite{lu:optimization} use a greedy algorithm to generate
query plans for multi-join queries that determines the order of each
join, but also the number of parallel joins and the processor assignment
to each join.
Ioannidis and Kang \cite{Ioannidis91} show that optimization for bushy
and left-deep trees is easier than optimizing left-deep trees alone.
Chen et al.~\cite{chen:segmented} proposed bushy trees with right-deep
pipelines, which they call segmented right-deep trees, for efficient
execution of multi-join queries. 
Query evaluation strategies for shared-nothing main-memory database
management systems have been examined in the context of the PRISMA/DB
system \cite{wils:parallel}.
Wilschut and Apers have identified synchronization costs as the 
performance bottleneck from parallelism \cite{wils:dataflow}.
Recently, Giceva et al.~\cite{alonso15} have studied how to allocate
resources and deploy query plans on multi-cores in the context of their
operator-centric SharedDB system.

Liu and Rundensteiner \cite{bin:revisiting} 
introduced segmented bushy query trees for the efficient evaluation of
multi-join queries.
Pipeline delays have also been studied in research by Deshpande and
Hellerstein \cite{desh:flow} who propose to reduce pipeline delays by
plan interleaving between plan fragments. 
Ahmed et al.~\cite{of_snowstorms_and_bushy_trees} have proposed a technique
to optimize bushy join trees for snowstorm queries in the Oracle
database system.

Machine learning techniques have also been explored to predict the
response time of a query.
IBM's DB2 introduced LEO, a learning optimizer, to adjust the query optimizer's
prediction based on the observed query cost~\cite{leo01,leo02}.
Zhang et al.~\mbox{\cite{comet05}} proposed a statistical learning
approach, COMET, to adapt to changing workloads for XML operations.
Duggan et al.~\mbox{\cite{dugganbal11}} used machine learning techniques
to predict the performance of concurrent workloads.
Akdere et al.~\mbox{\cite{learningmodel12icde}} introduced two learning-based
models, a plan-level model and an operator-level model, and contributed a
hybrid model to predict query performance.
Li et al.~\mbox{\cite{lisql12}} combined regression tree models and
scaling functions to estimate the resource consumption of individual
operators.

\begin{figure}
\centering
\includegraphics{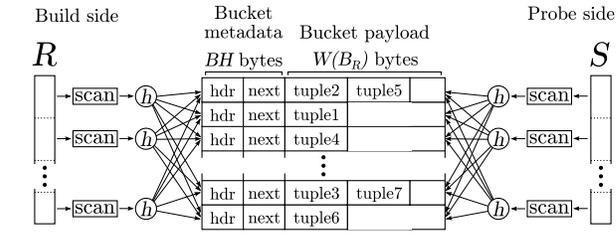}
\caption{
The non-partitioned hash join algorithm.
}
\label{fig:joinalgo}
\end{figure}

\subsubsection*{The non-partitioned in-memory hash join algorithm}
Our proposed memory I/O model captures the fundamental memory access
operations from the popular non-partitioned in-memory hash join
algorithm and the data structure layout optimization
by Balkesen et al.~\cite{cagri_hashjoin}. The main data structure is a
hash table that is shared by all threads (see Figure~\ref{fig:joinalgo}). 
The hash table is a contiguous array of buckets.
Each bucket contains metadata and the payload. 
Metadata are header information and a pointer to overflow buckets if
this bucket has spilled. 
The hash join is divided into two phases: the build phase and probe phase. 
During the build phase, an empty hash table is created and the build
table $R$ is divided into equi-sized parts which are assigned to
different threads.
All threads then scan their input sequentially, hash the join key of
every tuple, lock the bucket, write the tuple into the bucket, and
then unlock the bucket. 
The probe phase starts after the entire build table $R$ has been consumed.
In the probe phase, all threads sequentially scan the probe table $S$,
hash the join key of every tuple, and then join with the tuple in the
corresponding bucket if the join keys are equal.

\section{Modeling memory I/O} 
\label{sec:memaccess}

A fundamental challenge in building memory I/O models is the
pull-based nature of data movement throughout the memory hierarchy. 
In stark contrast to the disk I/O hierarchy where data movement is
explicitly initiated by the application, application control of memory I/O 
is limited to prefetch, flush and non-temporal access hints. 
Complicating matters further, memory loads and stores are not only
triggered when retrieving user data for query processing, but also when
manipulating local variables, accessing the call stack and co-ordinating
with other CPU cores. 

Determining the components of a memory I/O cost model is therefore a
delicate balancing act.
It is equally important to include sufficient detail to make accurate
predictions but also avoid adding too many parameters
that may exaggerate the effect of minor
fluctuations in the model inputs because of over-fitting.
The latter is particularly important for memory I/O cost models that
need to be generic to avoid obsolescence in light of the quick pace of
innovation in hardware.

\begin{table}
\centering
\begin{tabular}{cl}
\toprule
Mnemonic & \multicolumn{1}{c}{Access pattern} \\
\midrule
\sr & Read one cache line sequentially \\
\rr & Read one random cache line \\
\sw & Write one cache line sequentially \\
\rw & Write one random cache line \\
\end{tabular}
\caption{
Access patterns captured by the model.
}
\label{tbl:accesspatterns}
\end{table}

When constructing performance models for general algorithms
it is important to model every level
of the cache hierarchy.
The non-partitioned hash join algorithm, however, is 
cache-oblivious and memory-bound.
Therefore, our thesis is that 
when modeling a sequence of non-partitioned hash joins
\emph{the query response time is dominated
by the cost of accessing main memory} and not individual caches.

Not all access patterns to memory are equally expensive, however.
Sequential access patterns are more efficient because they leverage
prefetching and result in few TLB misses. 
Similarly, writes are more expensive than reads because they require
exclusive access to the cache line under the MESIF cache coherence
protocol, which may require cache line invalidations in other CPU cores.
(We outline the four access patterns in Table~\ref{tbl:accesspatterns}.)
We thus model the response time for processing query $Q$ as being
proportional to the number of operations $N(\textsf{A})$ 
weighted by the performance cost
$w_\textsf{A}$ for each access pattern $\textsf{A} \in \{\sr, \rr, \sw,
\rw\}$:

\vspace{-1em}
\begin{equation}
\label{eq:totalcost}
\begin{split} 
Time(Q) \; \propto \;
& \; w_\sr \cdot N(\sr) + w_\rr \cdot N(\rr) \\ 
&  + w_\sw \cdot N(\sw) + w_\rw \cdot N(\rw)
\end{split} 
\end{equation}

The performance cost $w_\textsf{A}$ for each access pattern \textsf{A}
is obtained experimentally for each system through a controlled
microbenchmark procedure, which is described in detail in 
Section~\ref{sec:bootstrapping}. 
The number of operations $N(\cdot)$ is calculated from our model in
Sections~\ref{sec:singleop} and \ref{sec:multijoin}. 
Section~\ref{sec:insight} summarizes key insights obtained by our model.

\subsection*{Modeling assumptions and limitations}

Our model assumes that memory is shared by all processing cores, it is
automatically kept coherent by the hardware and that data
transfer operations are performed at the granularity of a cache line.
We also assume that the database and the working memory for a query can fit
in memory and no paging to secondary storage occurs.
We only consider multi-join queries that are evaluated in parallel from
all threads using the non-partitioned hash-join algorithm by Balkesen et
al.~\cite{cagri_hashjoin}, which has been shown to offer competitive
performance while judiciously using memory \cite{blanas:memory}.

The fact that our proposed model does not account for any caching
or NUMA effects does not imply that architecture-awareness is not important.
On the contrary, the memory I/O cost model is an abstraction of a
well-designed query engine that is not compute-bound and will not
trigger redundant memory I/O during query processing. 
Inefficient implementations of the non-partitioned hash join will result
in higher memory traffic, which will not be captured by the model.
In particular, we assume that the hash table is striped on all available
NUMA nodes.

Finally, a limitation of join cost modeling is the assumption that the
cardinality of intermediate join results is known  or can be estimated
reasonably accurately. We acknowledge that cardinality estimation is not
a solved problem \cite{is_qo_a_solved_problem}.
Inaccuracies in estimating the cardinality or the skew of the build side are
particularly problematic. Improperly sized hash tables will result in
the creation and probing of overflow hash buckets. 
Some DBMSs will track excessive bucket overflows and pause the hash
table build process to resize the hash table and accommodate
more tuples. 
New join algorithms have recently been proposed to avoid 
populating the hash table with heavily-skewed data \cite{trackjoin14}.
Bucket overflows and hash table resizes are not accounted for by the
proposed model.

\subsection{Bootstrapping the model}
\label{sec:bootstrapping}

The model depends on the performance cost $w_\textsf{A}$ for each access
pattern $\textsf{A} \in \{\sr, \rr, \sw, \rw\}$ for accurate predictions. 
These general performance costs encapsulate (1) differences in the
underlying hardware of each system such as different CPU models or
different memory modules, (2) some hardware configuration options such
as memory timings and NUMA settings that can be changed in the BIOS, and
(3) OS-specific configuration options, such as the number and size of
huge TLB pages.
Bootstrapping consists of four specific microbenchmarks where all threads 
transfer data from memory such that only one specific
access pattern is triggered at a time. The microbenchmarks do not perform any
computation, so the derived performance costs can be thought as a
``worst-case'' scenario where all threads are waiting on memory I/O.

To derive the performance cost for the \sr and \sw patterns each thread
allocates a NUMA-local contiguous array and populates it with
random values. 
Each array is sized to a power of two such that as much of the available memory in the
system is used, but no paging to secondary storage occurs.
Every thread then concurrently reads its local array sequentially and
either accumulates an 8-byte value if the access pattern is \sr, or
accumulates and writes an 8-byte value back if the access
pattern is \sw.
(Notice that the \sw access pattern includes an explicit memory read.)

The \rr and \rw microbenchmarks use an array of the same size as for \sr
and \sw that is now shared between all threads. 
Before the benchmarking, each thread
first generates a random sequence of cache-aligned array offsets in
NUMA-local memory and waits on a barrier.
The benchmarking then starts, each thread accesses an entire cache line 
at every offset in the randomly generated sequence. If the access
pattern is \rr, every thread only accumulates values in an 8-byte local
variable while reading the cache line. If the access pattern is \rw,
every thread reads and writes the entire cache line in 8-byte units.

The performance cost $w_\textsf{A}$ is calculated by dividing the
execution time of the $\textsf{A} \in \{\sr, \rr, \sw, \rw\}$
microbenchmark with the number of cache lines that were accessed.

\begin{table} 
\centering
\begin{tabular}{cl}
\toprule 
Symbol          & \multicolumn{1}{c}{Meaning} \\
\midrule
$\textsf{A}$    & Access pattern, one of \sr, \rr, \sw, \rw.  \\ 
$w_\textsf{A} $ & Performance cost for access pattern \textsf{A}  \\ 
$N(\textsf{A})$ & Number of operations of access pattern \textsf{A} \\
$CL$            & Cache line size 	\\
$|R|$           & Cardinality of relation $R$ \\
$W(R)$          & Size of each tuple in relation $R$ \\
$B_R$           & Hash table on $R$ \\
$|B_R|$         & Number of hash buckets in $B_R$ \\
$W(B_R)$        & Hash bucket size in $B_R$ \\ 
$T$             & Tuples per bucket (ie. hash table load factor)\\ 
$BH$            & Size of metadata per hash bucket \\ 
$Q_{m,k}$       & Query subtree $k$ at depth $m$ \\
$I_{m,k}$       & Intermediate output of $Q_{m,k}$ (cf.~Figure \ref{fig:leftrighttree})\\
\end{tabular}
\caption{Notation.}
\label{tab:symbols}
\end{table}

\subsection{Modeling individual operators}
\label{sec:singleop}

In this section we analyze the memory access cost of individual
operations. The notation is summarized in Table~\ref{tab:symbols}.

\subsubsection*{Sequential scan}

All threads read data sequentially from memory during the scan. 
Scanning entails no random reads or memory writes. 
For a relation $ R $, the number of \sr memory operations is
determined by the size of the input relation and cache line size.
Therefore:

\begin{equation} 
N_{scan(R)}(\textsf{\sr}) = \left \lceil \frac{|R| \cdot W(R)}{CL}
\right \rceil 
\end{equation} 

\begin{equation} 
N_{scan(R)}(\textsf{\rr}) =  
N_{scan(R)}(\textsf{\sw}) = 
N_{scan(R)}(\textsf{\rw}) = 0
\end{equation}

\subsubsection*{Hash join build phase}

During the hash join build phase every thread concurrently reads tuples from the
build relation, hashes the tuple on the join key and inserts it into a
shared hash table. 
Hashing the join key only involves
computation, thus there will be no memory accesses. 
As a good hash function is designed to assign different keys into
different hash buckets, the hash buckets will be accessed in random
order.
Once the appropriate hash bucket has been identified, the thread needs
to latch the bucket before writing into it to prevent insertion races
with other threads.
The latch and other metadata information are stored in a bucket header
structure at the beginning of the bucket and have size $BH$. 
Acquiring the latch will result in a local modification to the cache
line. In the absence of any latch contention, releasing the latch
will modify the same cache line again.
Since the latch has likely already been brought into the cache after the
latch acquire operation, the latch release operation will hit the
same cache line and will not trigger additional memory accesses.
Therefore one \rw
memory accesses will occur per tuple in the build relation $R$:

\begin{equation} N_{build(R)}(\rr) = N_{build(R)}(\sr) = 0 \end{equation}

\begin{equation} N_{build(R)}(\rw) = |R| \end{equation}

\begin{figure}
\centering
\includegraphics[scale=0.8]{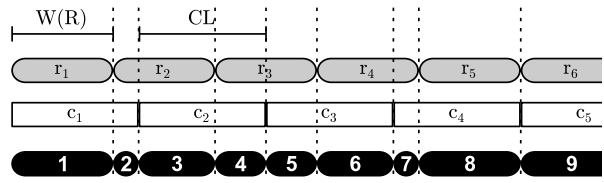}
\caption{
Modeling the cache line ($c_i$) writes in a hash bucket, when inserting
tuples $r_i$.
}
\label{fig:segments_and_cuts}
\end{figure}

Writing each $R$ tuple in the hash bucket will require additional writes.
The number of memory writes depends on
how many cache lines overlap with the modified memory region and the
write pattern.
Let us consider the insertion of $T$ tuples from relation $R$ (denoted
by $r_1, r_2, \ldots$) into the hash bucket that consists of the cache
lines $c_1, c_2, \ldots$ that is shown in Figure~\ref{fig:segments_and_cuts}. 
When the first tuple $r_1$ is inserted it fits entirely into $c_1$,
causing one memory write. 
When $r_2$ is inserted, it spans both the $c_1$ and the
$c_2$ cache lines and requires two memory writes. 
Similarly,
both $r_3$ and $r_4$ trigger two memory writes each because they span
cache lines, but inserting $r_5$ only requires one memory write
as it fits completely within the $c_4$ cache line.

We can derive a closed form solution for the memory writes that are
triggered by inserting $T$ tuples of width $W(R)$ into cache lines of
size $CL$ by calculating the boundaries of the cache lines and the
tuples, which are shown as vertical dashed lines in
Figure~\ref{fig:segments_and_cuts}.
We apply the inclusion/exclusion principle and first add the
tuple boundaries ($k_1$) and cache boundaries ($k_2$). We will then
subtract the boundaries that are both tuple and cache line
boundaries ($k_3$), such as the boundary between $r_5$ and $r_6$ 
in Figure~\ref{fig:segments_and_cuts}.
The $r_1, \ldots, r_T$ tuples have $k_1 = T-1$ boundaries and the
cache lines have $k_2 = \left\lfloor\frac{T \cdot W(R) - 1}{CL}\right\rfloor$ boundaries.
The common boundaries can be found in the locations that are both a
multiple of $W(R)$ and a multiple of $CL$. 
If $lcm(a, b)$ is the least common multiple of $a$ and $b$, 
these locations are precisely
at $1 \times lcm\big(W(R), CL\big)$, 
$2 \times lcm\big(W(R), CL\big)$, 
$\ldots$. 
Thus, the number of common boundaries is $k_3 = \left\lfloor\frac{T \cdot
W(R)-1}{lcm\big(W(R), CL\big)}\right\rfloor$. 
We conclude that the number of memory
accesses to insert $T$ tuples of size $W(R)$ into one hash bucket 
will be $\mathcal{M}(T, W(R)) = k_1 + k_2 - k_3 + 1$, which is:

\begin{equation}
\begin{split}
\mathcal{M}\big(T, &W(R)\big) = \\
&T
+ \left\lfloor\frac{T \cdot W(R) - 1}{CL}\right\rfloor
- \left\lfloor\frac{T \cdot W(R)-1}{lcm\big(W(R), CL\big)}\right\rfloor
\end{split}
\end{equation}

If the hash table $B_R$ on relation $R$ has $|B_R|$ hash buckets and each
bucket is sized to fit $\left\lceil\sfrac{|R|}{|B_R|}\right\rceil$ tuples,
the total number of \sw memory accesses is:

\begin{equation}
N_{build(R)}(\sw) = |B_R| \times
\mathcal{M}\Bigg(\left\lceil\frac{|R|}{|B_R|}\right\rceil, W(R) \Bigg)
\end{equation}

\subsubsection*{Hash join probe phase}

After the build phase completes, each thread will read tuples from the probe
relation $S$, hashes each tuple on the join key and compares the join
key of the tuple
from $S$
with the join keys of all the tuples in the corresponding hash bucket. 
No memory access is involved when evaluating the hash function. 
The join key will then need to be compared with every key in the hash
bucket. (If the hash bucket is empty, the hash bucket metadata will
still need to be retrieved to discover this.)
As the bucket is determined by the hash of the join key of the input tuple, the
first access to the hash bucket is \rr for every tuple in the probing
relation $S$:

\begin{equation} 
N_{probe(S)}(\rr) = |S| 
\end{equation}

All tuples in the hash bucket will be retrieved from memory for the join
key comparison using the \sr access pattern. 
The number of sequential reads is determined by the number of
additional cache lines the bucket occupies.
Each bucket in the hash table $B_R$ will contain $W(R) \cdot T + BH$
bytes of data, but the first $CL$ bytes have already been accounted for
in the \rr access pattern. 
Assuming that each hash bucket fits $T =
\left\lceil\sfrac{|R|}{|B_R|}\right\rceil$ tuples, let $\mathcal{L}(R)$ 
be the number of \sr memory accesses per probe: 

\begin{equation}
\mathcal{L}(R) 
= 
\left\lceil
\frac{
W(R) \cdot \left\lceil\frac{|R|}{|B_R|}\right\rceil + BH
}{
CL
}
\right\rceil - 1
\end{equation}

\begin{equation}
N_{probe(S)}(\sr) = |S| \times \mathcal{L}(R)
\end{equation}

The output of the join will be written to an output buffer of each
thread so as to be used by the next operator. 
Since the queue buffer is reused between operators, it can be sized to
fit in the cache and not generate any memory access when populating the
output buffer in a well-tuned system. 
Our model of the probe phase of the hash join therefore includes no \sw
or \rw operations:

\begin{equation}
N_{probe(S)}(\sw) = N_{probe(S)}(\rw) = 0
\end{equation}

\begin{figure}
\centering
\includegraphics{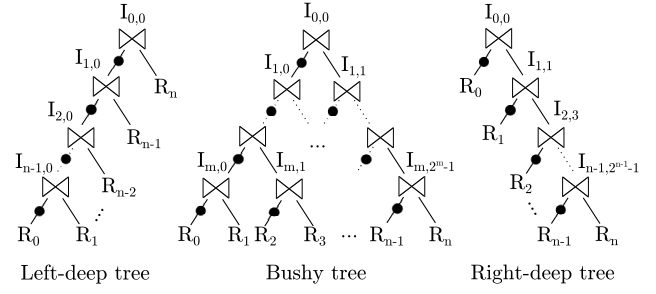}
\caption{Left-deep tree, bushy tree and right-deep tree. 
The output of a join is pipelined into the next operator.
The black dot denotes the build side, that is the join input that will be buffered in a hash
table before processing the probe side.}
\label{fig:leftrighttree}
\end{figure}

\subsection{Modeling multiple joins in a query tree}
\label{sec:multijoin}

The number of memory accesses for the subtree rooted at depth $i$ and
position $j$ (see the bushy tree in Figure~\ref{fig:leftrighttree})
is obtained by adding the memory accesses of the operator at depth $i$
and position $j$ with the memory accesses of the left and right subtrees
for each access pattern $\textsf{A} \in \{\sr, \rr, \sw, \rw\}$.
If $I_{i,j}$ denotes the intermediate result from the operator at depth
$i$ and position $j$, the number of memory accesses with pattern
\textsf{A} for the entire query is $N(\textsf{A})=N_{0,0}(\textsf{A})$,
where: 

\begin{equation}
N_{i,j}(\textsf{A}) =
\begin{cases}
N_{scan(R_k)}(\textsf{A}) & \hspace{-1cm}\text{if seq. scan,} \vspace{1em} \\
N_{build(I_{i+1,2j})}(\textsf{A}) + N_{probe(I_{i+1,2j+1})}(\textsf{A}) \\
+ N_{i+1,2j}(\textsf{A}) + N_{i+1,2j+1}(\textsf{A}) & \hspace{-1cm}\text{if hash join.}
\end{cases}
\label{eq:puttingitalltogether}
\end{equation}

Equation \ref{eq:puttingitalltogether} does not account for the cost of
materializing the output $I_{i,j}$. 
For intermediate results, we have experimentally found that this is not
a significant cost if the output buffer has been sized appropriately
\cite{monetdbx100} --- the buffer that stores the materialized output
batch for the next operator stays cached and does not trigger memory
accesses. (Note, however, that the cost of processing a larger intermediate
result is captured in the cardinality and width of $I_{i,j}$ when
accounting for the memory accesses of subsequent operations.)  
We validate this modeling choice in Section~\ref{sec:probe}.
The final join output $I_{0,0}$ is also materialized incrementally if it
is consumed by an aggregation or a top-$k$ operation. (This pattern, for
instance, occurs in every query in the TPC-H decision support
benchmark.)
If the entire query output is to be materialized, the model
can be trivially extended with additional \sw accesses to reflect the
materialization cost. This cost would be identical for all query plans.

\subsection{Insight on join evaluation orders}
\label{sec:insight}

This model can offer
valuable insights into the memory I/O requirements of different query plans.
We illustrate this by modeling the common case of a linear primary
key--foreign key join between $n+1$ relations.
Let that the multi-join operation be
$R_0 \Join R_1 \Join \cdots \Join R_{n-1} \Join R_n$, where 
$|R_i| \leq |R_{i+1}|$, for every $i \in [0,n-1]$, 
The join sequence for a left-deep evaluation strategy $Q_L$ is:
$$ Q_L = \bigg(\Big(\big(\left(R_0 \Join R_1\right) \Join \cdots\big) \Join
R_{n-1} \Big) \Join R_n \bigg)$$
While the right-deep evaluation strategy $Q_R$ is:
$$ Q_R = \bigg(R_0 \Join \Big(R_1 \Join \big(\cdots \Join \left(
R_{n-1} \Join R_n \right)\big)\Big)\bigg)$$

Let us consider the common
case of a join that produces as many tuples as the probe side, as is
commonly the case for primary key--foreign key joins. 
Closed formulas for the memory access cost of the right-deep query tree $Q_R$
that is shown in Figure~\ref{fig:leftrighttree} can be obtained by
setting $I_n = R_n$, as follows:

\begin{equation}
\begin{split}
N(\sr) 
&= \sum_{i=0}^{n}\left ( \left \lceil \frac{|R_i| \cdot
|W(R_i)|}{CL} \right \rceil \right )
+ \sum_{i=1}^{n}\left( |I_{i}| \times \mathcal{L}(R_{i-1}) \right)
\end{split}
\end{equation}

\begin{equation}
\begin{split}
N(\sw) 
&= \sum_{i=0}^{n-1}\left ( |B_{R_i}| \times 
\mathcal{M}\left ( \left \lceil \frac{|R_i|}{|B_{R_i}|} \right \rceil, W(R_i) \right ) \right )
\end{split}
\end{equation}

\begin{equation} 
N(\rr) 
= \sum_{i=1}^{n}|I_i| 
\qquad \qquad
N(\rw) 
= \sum_{i=0}^{n-1}|R_i|
\end{equation}

The cost for a left-deep tree $Q_L$, with $I_n = R_0$ as shown in
Figure~\ref{fig:leftrighttree}, is:

\begin{equation}
N(\sr) =
\sum_{i=0}^{n}\left ( \left \lceil \frac{|R_i| \cdot
|W(R_i)|}{CL} \right \rceil \right ) 
+ \sum_{i=1}^{n}\left ( |R_i| \times \mathcal{L}(I_{n-i+1}) \right)
\end{equation}

\begin{equation}
N(\sw)
= \sum_{i=1}^{n}\left ( |B_{I_i}| \times 
\mathcal{M}\left ( \left \lceil \frac{|I_i|}{|B_{I_i}|} \right \rceil, W(I_i) \right ) \right ) \\
\end{equation}

\begin{equation} 
N(\rr) 
= \sum_{i=1}^{n}|R_i| 
\qquad \qquad
N(\rw) 
= \sum_{i=1}^{n}|I_i|
\end{equation}

\textbf{Insight 1: The left-deep ($Q_L$) and right-deep ($Q_R$) evaluation
strategy perform the same number and type of memory writes.}
$N(\rw)$ and $N(\sw)$ are identical for the two query trees, because
the cardinality of the $k$-th intermediate output is $|I_k| = |R_{n-k}|$
for the left-deep strategy $Q_L$, and $|I_k| = |R_n|$ for the right-deep
strategy $Q_R$. 

\textbf{Insight 2: The left-deep
($Q_L$) evaluation strategy performs $n|R_n| - \sum_{i=1}^{n}|R_i|$ fewer
\rr memory accesses than the right-deep ($Q_R$) strategy.}
The \rr accesses for the left-deep strategy are $N_L(\rr) =
\sum_{i=1}^{n}|R_i|$, while the \rr accesses are $N_R(\rr) = n|R_n|$ for
the right-deep strategy.
$N(\sr)$ is identical in both strategies if $\mathcal{L}(\cdot) = 0$,
namely if the tuples are thinner than the last-level cache line.
This is commonly the case due to projection push down or columnar 
storage.

\subsection{Possible extensions to the model}
The proposed model accounts for wide tuples but assumes that the tuple width
is fixed per input relation. This may not be accurate if some
attributes represent variable length fields such as VARCHARs, BLOBs or
nested structures like JSON objects. 
The model could crudely be applied by using the \emph{average} tuple
size, but accurate results would require a complete histogram of the
attribute size per join key. A research challenge is to devise
techniques to collect such metadata information efficiently.

Another promising direction is to extend the model for
distributed in-memory joins. 
We foresee two challenges in generalizing our model to distributed
in-memory hash joins: modeling data placement and locality and modeling
the cost of remote data transfers. 
Our model currently does not consider data placement.
It seems unlikely that a model
that does not account for locality can remain accurate when some
accesses become remote and thus an order of magnitude more expensive
than what is currently modeled. 
Regarding modeling the cost of remote data transfers, Barthels et
al.~\cite{rackjoin15} have proposed a performance model to estimate the
cost of transferring data using RDMA primitives for join processing.

Hash tables are widely used for storing data for in-memory caching or
in-memory transaction processing. Modeling the cost of each access 
is a first step towards understanding the performance
characteristics of complex operations (such as transactions) that
consist of a series of reads and writes on a hash table.
We speculate that the main limitation of our model when applied to ad-hoc
operations is to account for the effect of concurrent writes and reads.
A promising research direction is to construct contention-aware
performance models for in-memory data processing.

\begin{table}[b]
\small
\newcommand{\mc}[3]{\multicolumn{#1}{#2}{#3}}
\centering
\begin{tabular}{ccccc}
\toprule
& \mc{4}{c}{Cost per access pattern} \\
\cmidrule(rl){2-5} 
Vendor, CPU or instance type
& $w_\sr$   & $w_\rr$   & $w_\sw$   & $w_\rw$   \\ 
\midrule
Intel, 2$\times$E5-2695v2  & 1.00 & 3.79 & 5.03 & 6.25 \\
AMD, 2$\times$Opteron 6172 & 1.00 & 6.44 & 1.88 & 8.42  \\
Amazon EC2, c4.4xlarge     & 1.00 & 6.81 & 5.21 & 13.86 \\
\end{tabular}
\caption{
Measured cost per access pattern for all systems.
(See Section~\ref{sec:bootstrapping} for details on the bootstrapping
process).
}
\label{tbl:weights}
\end{table}

\begin{table*}[t]
\newcommand{\mc}[3]{\multicolumn{#1}{#2}{#3}}
\newcommand{\oc}[1]{\multicolumn{1}{c}{#1}}
\small
\centering
\begin{tabular}{crrrrrrrrrrrr} 
\toprule
& \mc{5}{c}{Predicted memory I/O}	 
& \mc{4}{c}{Observed memory I/O}
& \mc{3}{c}{Response time slowdown} \\
\cmidrule(rl){2-6}
\cmidrule(rl){7-10}
\cmidrule(rl){11-13}
\multirow{2}{3em}[-0.5em]{\centering Load factor}
& \oc{\rw}	 & \oc{\sw}
& \multirow{2}{3em}[-0.5em]{\centering Total writes}
& \oc{\sr}
& \multirow{2}{3em}[-0.5em]{\centering Total reads}
& \mc{2}{c}{Writes}
& \mc{2}{c}{Reads}
& \multirow{2}{3em}[-0.25em]{\centering Pred.}
& \multirow{2}{3em}[-0.25em]{\centering Observ.}
& \multirow{2}{3em}[-0.25em]{\centering {\bf Error}}
\\
\cmidrule(rl){2-2}
\cmidrule(rl){3-3}
\cmidrule(rl){5-5}
\cmidrule(rl){7-8}
\cmidrule(rl){9-10}
&{\scriptsize Build($R$)}	
&{\scriptsize Build($R$)}
&
&{\scriptsize Scan($R$)}
&
& Total
& {\bf Error}
& Total
& {\bf Error}
\\
\midrule                            
1.0	&   512 &     0 &   512	 &   128 &   640 &   505 & {\bf 1.4\%} &   634 & {\bf  0.9\%} &     \oc{---}   &    \oc{---}    &   \oc{---}   \\
2.0	& 1,024 &     0 & 1,024	 &   256 & 1,280 & 1,019 & {\bf 0.5\%} & 1,277 & {\bf  0.2\%} & $ 2.00 \times$ & $ 1.97 \times$ & {\bf  1.5\%} \\
3.0	& 1,536 &     0 & 1,536	 &   384 & 1,920 & 1,531 & {\bf 0.3\%} & 1,921 & {\bf -0.1\%} & $ 3.00 \times$ & $ 2.94 \times$ & {\bf  2.0\%} \\
4.0	& 2,048 &   512 & 2,560	 &   512 & 3,072 & 2,555 & {\bf 0.2\%} & 3,081 & {\bf -0.3\%} & $ 4.77 \times$ & $ 4.77 \times$ & {\bf  0.0\%} \\
5.0	& 2,560 & 1,024 & 3,584	 &   640 & 4,224 & 3,580 & {\bf 0.1\%} & 4,234 & {\bf -0.2\%} & $ 6.55 \times$ & $ 6.56 \times$ & {\bf -0.2\%} \\
6.0	& 3,072 & 1,536 & 4,608	 &   768 & 5,376 & 4,605 & {\bf 0.1\%} & 5,388 & {\bf -0.2\%} & $ 8.32 \times$ & $ 8.36 \times$ & {\bf -0.5\%} \\
7.0	& 3,584 & 2,048 & 5,632	 &   896 & 6,528 & 5,629 & {\bf 0.1\%} & 6,548 & {\bf -0.3\%} & $10.09 \times$ & $10.14 \times$ & {\bf -0.5\%} \\
8.0	& 4,096 & 2,560 & 6,656	 & 1,024 & 7,680 & 6,656 & {\bf 0.0\%} & 7,701 & {\bf -0.3\%} & $11.87 \times$ & $11.98 \times$ & {\bf -0.9\%} \\
\end{tabular}
\caption{Model accuracy for different hash table load factors. I/O events in millions.}
\label{tab:buildmodel}
\end{table*}

\begin{table*}[t]
\newcommand{\mc}[3]{\multicolumn{#1}{#2}{#3}}
\newcommand{\oc}[1]{\multicolumn{1}{c}{#1}}
\small
\centering
\begin{tabular}{ccrrrrrrrrrrr}
\toprule
&
&
& \mc{5}{c}{Predicted memory I/O}	 
& \mc{2}{c}{Observed I/O}
& \mc{3}{c}{Response time slowdown} 
\\
\cmidrule(rl){4-8}
\cmidrule(rl){9-10}
\cmidrule(rl){11-13}
\multirow{2}{*}[-0.5em]{\centering \large ${|S|}\over{|R|}$}
& \multirow{2}{2.75em}[-0.5em]{\centering Join \hspace*{-0.8em}selectivity} 
& \multirow{2}{4em}[-0.5em]{\centering $|S \Join R|$ {\scriptsize (millions)}}
& \oc{\rw}
& \oc{\rr}
& \mc{2}{c}{\sr}
& \multirow{2}{3em}[-0.25em]{\centering Total}
& \multirow{2}{3em}[-0.25em]{\centering Total}
& \multirow{2}{3em}[-0.25em]{\centering {\bf Error}}
& \multirow{2}{3em}[-0.25em]{\centering Pred.}
& \multirow{2}{3em}[-0.25em]{\centering Observ.}
& \multirow{2}{3em}[-0.25em]{\centering {\bf Error}}
\\
\cmidrule(rl){4-2}
\cmidrule(rl){5-5}
\cmidrule(rl){6-7}
& 
& 
& {\scriptsize Build($R$)}
& {\scriptsize Probe($S$)}
& {\scriptsize Scan($R$)}
& {\scriptsize Scan($S$)}
&
\\
\midrule
$4 \times$ &  25\% &   512 & 512 & 2,048 & 128 &   512 & 3,200 & 3,136 & {\bf  2.0\%} & $2.15 \times$ & $2.04 \times$ & {\bf 5.4\%} \\
$3 \times$ &  33\% &   512 & 512 & 1,536 & 128 &   384 & 2,560 & 2,625 & {\bf -2.5\%} & $1.77 \times$ & $1.71 \times$ & {\bf 3.5\%} \\
$2 \times$ &  50\% &   512 & 512 & 1,024 & 128 &   256 & 1,920 & 1,957 & {\bf -1.9\%} & $1.38 \times$ & $1.37 \times$ & {\bf 0.7\%} \\
\cmidrule(rl){1-13}
$1 \times$ & 100\% &   512 & 512 &   512 & 128 &   128 & 1,280 & 1,299 & {\bf -1.5\%} &   ---         &   ---         &        ---  \\
\cmidrule(rl){1-13}
$2 \times$ & 100\% & 1,024 & 512 & 1,024 & 128 &   256 & 1,920 & 1,876 & {\bf  2.3\%} & $1.38 \times$ & $1.37 \times$ & {\bf 0.7\%} \\
$4 \times$ & 100\% & 2,048 & 512 & 2,048 & 128 &   512 & 3,200 & 3,270 & {\bf -2.1\%} & $2.15 \times$ & $2.10 \times$ & {\bf 2.4\%} \\
$6 \times$ & 100\% & 3,072 & 512 & 3,072 & 128 &   768 & 4,480 & 4,583 & {\bf -2.2\%} & $2.92 \times$ & $2.84 \times$ & {\bf 2.8\%} \\
$8 \times$ & 100\% & 4,096 & 512 & 4,096 & 128 & 1,024 & 5,760 & 5,907 & {\bf -2.5\%} & $3.68 \times$ & $3.56 \times$ & {\bf 3.4\%} \\
\end{tabular}
\caption{Model accuracy for larger probe relations and different join selectivities. I/O events in millions.}
\label{tab:probemodel}
\vspace{-1em}
\end{table*}

\section{Experimental evaluation}
\label{sec:results}

This section describes the experimental evaluation and validation of our
model.
Sections~\ref{sec:build} and \ref{sec:probe} focus on a single join and evaluate the accuracy
of the model in predicting the memory I/O activity. We validate that I/O
activity is linearly correlated with query response time, and we find
that the impact of incrementally materializing intermediate results
to the next pipeline stage is negligible.
Section~\ref{sec:parvsseq} evaluates how parallel subtree evaluation affects
performance, and validates that sequential subtree evaluation results in
similar query response times to the fastest parallel evaluation strategy
but with less memory pressure.

We then turn our attention to multi-join queries, and we exhaustively
compare all possible multi-join query plans for a join with four
input relations.
Section~\ref{sec:diskmodel} discusses how our simple memory I/O model compares
with a linear disk-based I/O model, such as the Haas et al.~model
\cite{HaasS97:seeking}, and whether parameter tuning of disk-based models
is sufficient to predict the response time of queries on memory-resident
data.
Section~\ref{sec:amd} highlights how our model adapts to different
hardware and can accurately predict response time ``drifts'' of
particular query plans as the memory hierarchy evolves.
The remainder of the section evaluates the sensitivity of our model to
changes in the database size (Section~\ref{sec:varysize}), 
join skew (Section~\ref{sec:varyskew}), input cardinality ratio 
(Section~\ref{sec:varyratio}) and 
join selectivity (Section~\ref{sec:varyselectivity}). Finally
Section~\ref{sec:morejoin} evaluates the predictive power of our model in
forecasting the cost of deeper multi-join pipelines.

We have extended Pythia \cite{pythia}, a prototype open-source in-memory
query engine to support multi-way joins and complex query pipelines.
We started with the optimized non-partitioned hash join implementation
of Balkesen et al.~\cite{cagri_hashjoin} to create a hash join operator
that can be placed in complex query pipelines using a parallel,
vectorized, pull-based execution model \cite{monetdbx100, Volcano}.
Building on prior work \cite{chen_ailamaki_prefetching}, we prefetch
hash buckets before they are accessed by ``peeking ahead'' within the
batch of input tuples.
We have built a custom memory allocator to use huge pages for
memory requests larger than 1 GB. 
By tightly controlling memory allocation we can enforce NUMA placement,
support pre-allocation during query planning time, and
precisely monitor the memory consumption of a query during processing. 

\subsection{Hardware setup and methodology}
\label{sec:hardware}
We use three systems for our experimental evaluation.
The first system (which we will refer to as ``Intel'') is a server with two NUMA
nodes and 256 GB of DDR3 memory, with two Intel Xeon E5-2695v2
12-core processors. 
Each core in an E5-2695v2 processor shares the 30 MB L3 cache.
We have enabled hyper-threading, so the operating system can
simultaneously execute 48 threads. 
The second system (which we will refer to as ``AMD'') is a server with
four NUMA nodes and 32 GB of memory, with two
AMD Opteron 6172 12-core processors (in total, 24 concurrent threads). 
Each core in an Opteron 6172 processor shares
the 12 MB L3 cache. 

In a cloud environment, the precise hardware specifications may not
be known before the VM is instantiated.
An advantage of the proposed cache-oblivious model is its adaptability to 
hardware with unknown specifications.
The third system (which we refer to as ``EC2'') is
an Amazon EC2 instance of type \texttt{c4.4xlarge}. The instance
has 30 GB of memory and 16 vCPUs.
To control for cyclical variations in performance, 
we bid for a single spot EC2 instance around 10am
every morning (US Eastern time) for 7 consecutive
days. 
The observed day-to-day variation in performance is
less than 2\% so we only report averages over all
seven days for all EC2 experiments.

The measured performance cost $w_\textsf{A}$ for each access pattern
\textsf{A} among \sr, \rr, \sw and \rw for the three systems is shown in
Table~\ref{tbl:weights}. (Please refer to
Section~\ref{sec:bootstrapping} for details on the bootstrapping
process.)
We use the built-in ``perf'' tool
of the Linux kernel
to tap into hardware counters of performance events related to memory accesses.
To minimize reporting overheads from ``perf'' 
we only inspect these counters after
the completion of the initialization, the build and the probe phases
(that is, three times per join operation). 
Every experiment is repeated at least ten times, and we report the
average.
We assume that a memory manager preallocates memory when the
DBMS starts, and queries can quickly reuse pre-allocated memory. 
Thus, memory allocation time is not included in the reported query response
time.

\subsection{Validating modeling assumptions}
\label{sec:validate}

This section validates key modeling assumptions using the Intel system
that is described in Section~\ref{sec:hardware}. We first dissect a single hash
join to evaluate how accurately we model memory I/O traffic and
whether memory traffic is a good predictor of the join execution time.
We then turn our attention to how parallel evaluation of distinct build
subtrees of a query plan affects performance.

\subsubsection{Hash join build phase}
\label{sec:build}
The first experiment evaluates the accuracy of the model in capturing
the memory I/O activity and predicting the completion time of the hash join
build phase.
We create one relation $R$ with two attributes.
The first attribute is an 8-byte integer which ranges from 1 to the
cardinality of relation $|R|$, and the second attribute is a random 8-byte
integer.
Narrow tuples are
commonly encountered in column-stores and have been extensively used for
experimental evaluation in prior work \cite{cagri_hashjoin}.
The input relation has been randomly permuted so that the keys are in
random order, and is striped among both NUMA nodes of our system.
For this experiment we fix the number of hash buckets $|B|$ to 512
million and use 16 bytes per bucket as the bucket header $BH$; the first
cache line can therefore hold 3 tuples. 

The experimental results and the model predictions for building a hash
table are shown in Table~\ref{tab:buildmodel}.
We vary the cardinality of the build relation $R$ from 512 million $(1 \times)$
to 4 billion $(8\times)$ tuples. 
When the build relation size varies from $1\times$ to $3\times$, all
tuples will be in the same cache line with bucket header, and $N(\sw)=0$. 
When the build relation size reaches or exceeds $4\times$, tuples will
be inserted in the next available cache line and will cause \sw traffic.
When comparing the model prediction with the observed I/O activity, 
we notice that the read and the write prediction is very accurate and
commonly deviates less than 1\% from the observed memory traffic.
When comparing the 
model prediction 
with the experimentally observed execution time, we observe that  
the memory I/O activity is an accurate predictor of the execution time
of the build phase, as it commonly deviates less than 2\% from the
observed time.

\subsubsection{Hash join probe phase}
\label{sec:probe}
We now evaluate the accuracy of our model in capturing the hash join probe
phase.
We use the same build relation $R$ as in Section~\ref{sec:build} and fix the
cardinality of $R$ to 512 million tuples ($1\times$). 
As shown in Table~\ref{tab:probemodel}, the cardinality of the probe
relation $|S|$ varies from $1 \times 512M$ to $8 \times
512M$, and the probability that an $S$ tuple finds a matching $R$ tuple
in the hash table varies from 25\% to 100\%.
We observe that the model's memory activity prediction is accurate and
within 3\% of the observed memory read traffic. Memory I/O activity is
again linearly correlated with query response time. In fact, the model's
prediction slightly overpredicts the query response time. 
Finally, the cardinality of the join output $|S \Join R|$ does not
noticeably affect the memory I/O activity and the response time, because the
join output is materialized incrementally in the same output buffer for
pipelined operations. 
This validates our design choice to not explicitly account for the output
materialization cost of intermediate results.

\begin{figure}
\centering
\subfigure[
$k$ threads build the $J_1$ hash table, while the remaining
threads build the $J_2$ hash table.
All threads participate in the probe pipeline.
]{
\includegraphics{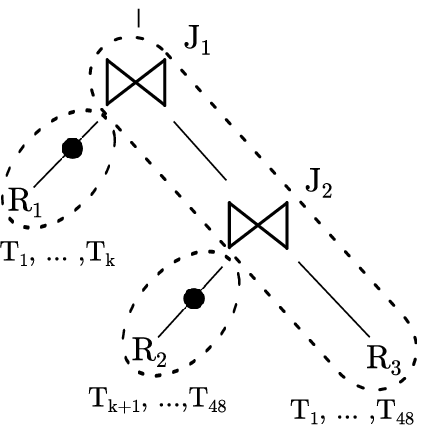} 
\label{fig:parallel.sequential.qp}
}
\hspace{1em}
\subfigure[Response time as the number of threads deviates from the optimal        
allocation.]{
\includegraphics[scale=0.85]{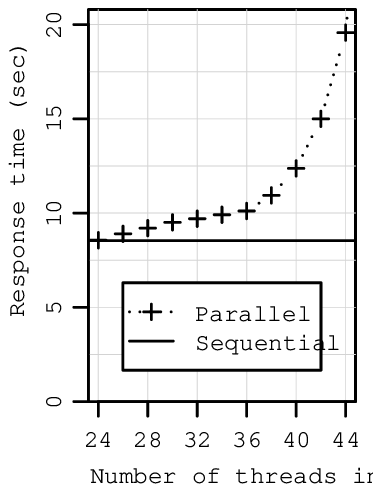}
\label{fig:parallel.sequential.rt}
}
\caption{
Exploring sequential or parallel build subtree evaluation for right-deep query
trees.
} 
\label{fig:parallel} 
\end{figure}

\subsubsection{Parallel vs. sequential build subtree evaluation}
\label{sec:parvsseq}
Schneider and DeWitt \cite{dono:tradeoff} argued that one aspect of superiority
of right deep trees over left deep trees is their potential for exploiting
parallelism in their independent build subtrees.
In this section, we study the performance of utilizing
this inter-subtree parallelism for memory-resident datasets. 
We create three relations that contain 1 billion tuples
each, and evaluate the query plan shown in
Figure~\ref{fig:parallel.sequential.qp}.
Every probe tuple matches one build tuple in the $J_1$ and $J_2$ joins. 
We execute the query plan both sequentially and in parallel.
When we execute the query sequentially we use all available threads to
build $R_1$ first, and then move on to build $R_2$ after completing
the build of $R_1$. 
When we execute the
query in parallel, the threads are divided
into two groups: $k$ threads build a hash table on $R_1$ while the other
$48-k$ concurrently build a hash table on $R_2$.
All threads participate in the probe phase for both evaluation strategies. 

The response time of the two evaluation strategies is shown in
Figure~\ref{fig:parallel.sequential.rt} on the vertical axis, as the number
of threads assigned to $R_1$ changes on the horizontal axis. 
Because the two relations have equal size, the optimal allocation is
$k=24$ which divides the threads into two equally-sized groups.
Performance will degrade if the optimal thread assignment is 
not predicted correctly by the query optimizer when processing
more complex query subtrees.
However, because the communication costs of a main-memory environment are
negligible compared to shared-nothing architectures, the performance of
building the hash tables on $R_1$ and $R_2$ sequentially is equal to the
performance of parallel evaluation with the best thread allocation.

An advantage of sequential evaluation is the memory consumption of
the query. 
When building
sequentially, 64GB of memory is allocated in the beginning, which
corresponds to the memory requirement for building $R_1$. The memory 
consumption further increases by 64GB (the amount of memory needed for
building $R_2$) later in the query, and remains constant. 
When building in parallel, 128GB of memory is needed upfront to start the
query, if both $R_1$ and $R_2$ are to be built concurrently.
Although the performance and the peak memory consumption is the same,
the sequential evaluation strategy imposes less pressure to the memory
manager and the DBMS because it gradually reaches its peak memory
allocation.
In light of these findings, we process
independent build subtrees sequentially in the experiments that follow.

\begin{figure}
\centering 
\includegraphics[scale=0.95]{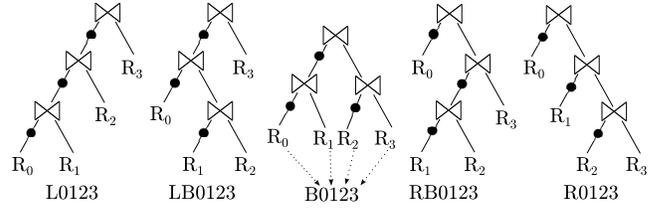} 
\caption{
Query plan notation for the exhaustive 4-relation join experiment. The
letters denote the tree shape, followed by the input relations when reading
from left to right.
} 
\label{fig:relation4queryplan} 
\end{figure}

\subsection{Evaluation with multi-join query plans}
\label{sec:eval:multijoinqp}

\label{sec:workload}

In order to exhaustively evaluate our model's prediction for multi-join query
plans, we generate a synthetic database with four relations $R_0$, $R_1$, $R_2$
and $R_3$. All relations contain two long integer attributes \texttt{a} and
\texttt{b}, and the tuple size is 16 bytes. 
The attribute \texttt{a} ranges from 1 to the cardinality of the relation and
the values are randomly distributed. 
We evaluate the following SQL query:

\vspace{1em}
\noindent
\texttt{
\hspace*{0.9em} SELECT SUM($R_0$.a + $R_3$.b) \\
\hspace*{1.5em} FROM $R_0$, $R_1$, $R_2$, $R_3$ \\
\hspace*{1.5em} WHERE $R_0$.b = $R_1$.a,  $R_1$.b = $R_2$.a, $R_2$.b = $R_3$.a
}
\vspace{1em} 

This synthetic database with 4 relations allows us to explore the five query
plan shapes which are shown in Figure~\ref{fig:relation4queryplan},
namely the left-deep (L), left-bushy (LB), bushy (B),
right-bushy (RB), and right-deep (R) tree.
We use ``TXXXX'' to refer to a query plan, where ``T'' is the query type,
followed by a sequence of numbers denoting the input relations when
reading from
left to right. 
By focusing on a database with 4 relations, we can systematically explore the
entire query plan space, while keeping the number of query plan candidates
within a reasonable limit for presentation purposes.
(We relax this constraint in Section~\ref{sec:morejoin} where we evaluate the
model with query plans that involve more than 3 joins.)

\begin{figure*}
\centering
\subfigure[PostgreSQL disk I/O cost model, $r_p$=0.740, $r_s$=0.724]{
\label{fig:modelseq}
\includegraphics[width=0.49\textwidth]{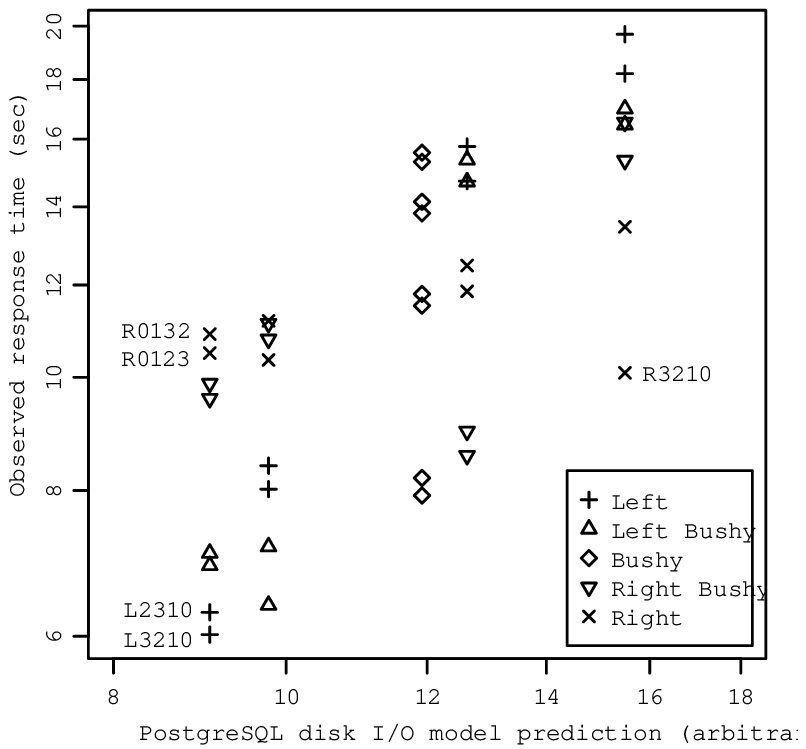}}
\subfigure[Our model on the Intel system, $r_p$=0.970, $r_s$=0.982.]{
\label{fig:modelour}
\includegraphics[width=0.49\textwidth]{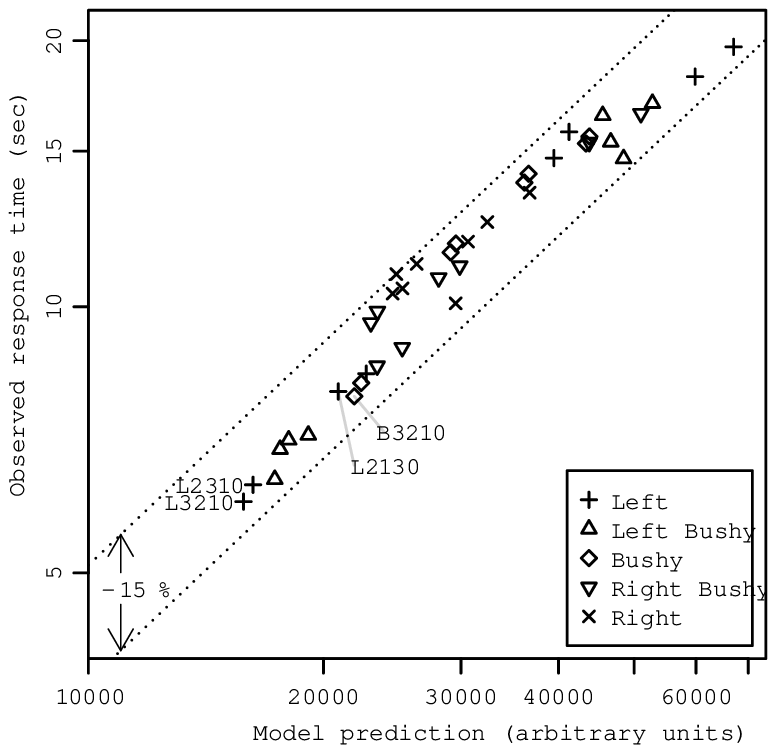}}
\subfigure[Our model on the AMD system, $r_p$=0.881, $r_s$=0.903.]{
\label{fig:amd}
\includegraphics[width=0.49\textwidth]{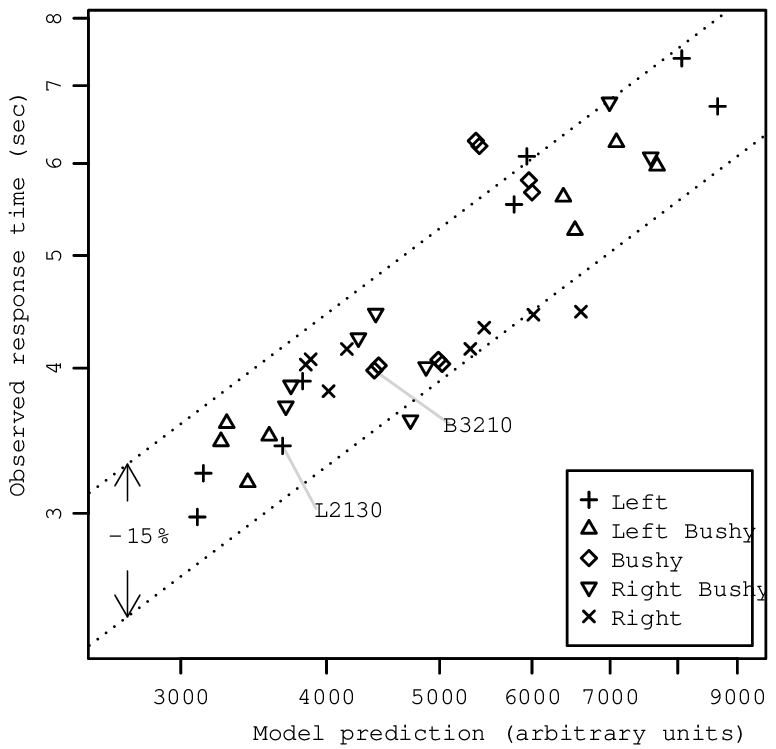}}
\subfigure[Our model on the EC2 instance, $r_p$=0.939, $r_s$=0.989.]{
\label{fig:ec2}
\includegraphics[width=0.49\textwidth]{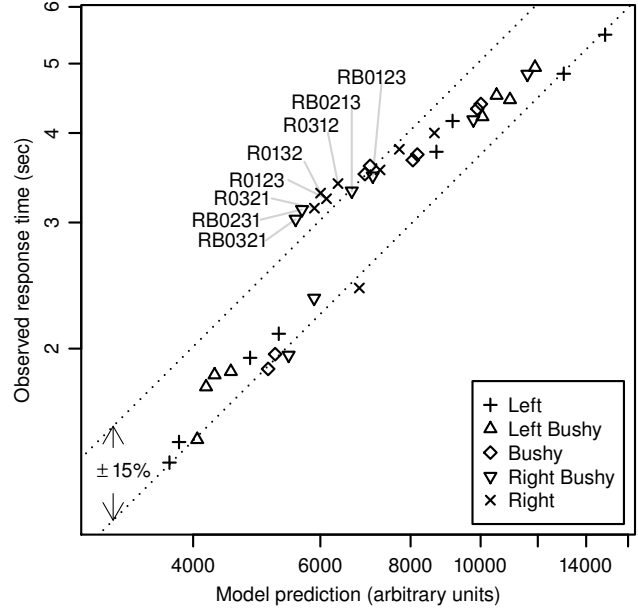}}
\caption{Observed query response time vs. predicted cost from the PostgreSQL
I/O cost model with least-squares fit, and our model for the Intel, AMD
and EC2 systems.
$r_p$ denotes the Pearson correlation coefficient, and $r_s$ the
Spearman correlation coefficient.} 
\label{fig:modelcomp}
\end{figure*}

\subsubsection{Comparison with a popular disk-based model}
\label{sec:diskmodel}

Do we need a new cost model for memory I/O, or is adjusting the weights
of existing linear disk-based models sufficient?
We answer this question by comparing the accuracy of our model with the
PostgreSQL 9.3 disk I/O model which
mimics the Haas et al.~\cite{HaasS97:seeking} model:
\begin{equation} 
	Cost(Q) \propto n_s \cdot c_s + n_r \cdot c_r
\end{equation}
The number of I/O operations $n_s$ and $n_r$ for a given query plan are
predicted by the query optimizer and is a function of the cardinalities of the
join inputs and the join output. 
The cost weights $c_s$ and $c_r$ are user-adjustable parameters. 

We exhaustively explore all viable multi-join query plans with 4
relations for the query that is shown in Section~\ref{sec:workload}. We set the relation
cardinality
of $R_0$, $R_1$, $R_2$, $R_3$ to 2 billion, 512 million, 128 million and 32
million tuples respectively. The join ratio is 1:4, that is for every value of
$R_{k+1}.a$, there are 4 tuples with matching $R_{k}.b$ values.
We force PostgreSQL to use a specific join order and explicitly set the 
cardinality estimates to the actual join cardinality instead of relying on the
prediction by the query optimizer.
We obtain $n_s$ by setting $w_s$ to 1 and $w_r$ to 0, and we report the total
cost for the query plan, as shown by PostgreSQL's \texttt{EXPLAIN} command. 
(We do likewise to find $n_r$.)
We finally perform linear regression to find the \emph{optimal} non-negative costs to
minimize the error between the cost prediction and the observed execution time.

Figure~\ref{fig:modelseq} shows the prediction of the
PostgreSQL disk I/O model on the horizontal axis (after parameter fitting
using linear regression) and the observed response time on the Intel
system on the vertical axis. 
Even with optimal cost weights, the
predictions of the PostgreSQL disk I/O model are clustered in five points
(shown as vertical bands) on Figure~\ref{fig:modelseq}.
PostgreSQL gives the same cost
prediction for plans with very different execution times: it predicts
that the best two plans (``L3210''
and ``L2310'') will have the same response time as the 
``R0132'' and ``R0123'' plans which are in reality twice as expensive. 
Making matters worse, ``R3210'' is the optimal right-deep query plan but is
predicted to be twice as expensive as the ``R0123'' query plan.

In comparison, Figure~\ref{fig:modelour} shows the prediction from our model on
the Intel system, which successfully predicts the cheapest plans
``L3210'' and ``L2310''. In addition, all observed response times lie within a narrow
$\pm 15\%$ band of their prediction.
A disk I/O model that only considers the size of the data accessed
and whether this access is sequential or random cannot
account for the access patterns that arise during an in-memory hash join.

\subsubsection{Model adaptability to different hardware}
\label{sec:amd}

In this section, we test the model's accuracy on the different systems
that were described in Section~\ref{sec:hardware}.
Because of limited memory size, we set the cardinality
of $R_0$, $R_1$, $R_2$, $R_3$ to be 256 million, 64 million, 16 million,
and 4 million tuples respectively. The join ratio is set to be 1:4 as in
Section~\ref{sec:diskmodel}, and the query is the 4-relation join
described in Section~\ref{sec:workload}.
The predictions are more ``noisy'' due to caching effects that arise
because of the smaller database size --- for instance, $R_3$ can be
cached in its entirety in all systems. (We explore the model sensitivity
to the database size in Section~\ref{sec:varysize}.)

Figure~\ref{fig:amd} compares the model's cost prediction with the
observed query response time on the AMD system.
Our model remains accurate as only a few of the 40
query plans fall outside the $\pm 15\%$ band.
The proposed model also adapts to the hardware architecture of the
underlying platform.
Observe the ``L2130'' and ``B3210'' query plans on the Intel system
at Figure~\ref{fig:modelour}. Both plans have been accurately predicted
to have comparable performance. Yet for the AMD system
at Figure~\ref{fig:amd}, the model accurately predicts that the
``B3210''
query plan is $15\%$ more expensive than the ``L2130'' plan.
This difference can be attributed to the proportion of the \rr activity
of the two plans, which is $76\%$ for the ``B3210'' plan but only $52\%$
for the ``L2130'' plan.
As the $w_\rr$ weight changes from 3.79 on the Intel system to 6.44 on
the AMD system (cf. Table~\ref{tbl:weights}), our model captures this
performance difference
and
adjusts the cost predictions for all plans accordingly.

We also test our model on the Amazon EC2 instance using the same dataset
as for the AMD system and show the results in Figure \ref{fig:ec2}.
The EC2 results are similar to the results from the Intel system,
except for all the plans with prefixes ``RB0'' and ``R0'' that
construct a hash table on $R_0$. Our model predicts that these plans are
much faster than their actual response time.
A closer look at the performance counters reveals that the discrepancy
is due to the cost to translate virtual to physical memory addresses
when
accessing this large hash table. The Intel system was configured with
1GB huge pages that Pythia can readily use, while the EC2 instance had
no
huge page support. Modeling the hierarchical nature of the page table
could improve the accuracy of the model
when huge pages are not enabled or not supported.

We now adopt more
concise metrics to quantify model accuracy.
Prior work \cite{gu12:testing, wu14:uncertain} has used correlation
metrics
to measure the accuracy of the model prediction.
The Pearson correlation coefficient $r_p$ is
used to measure the linear
correlation between two variables (namely, the prediction and the
actual query response time).
A deficiency of the Pearson coefficient is that it can
assign low coefficient scores to models that correctly
predict the relative ordering of different query plans, but the
correlation with the observed response time is non-linear.
The Spearman correlation coefficient $r_s$ accommodates non-linear
models by measuring the linear correlation between the \emph{ranks} of
different query plans.
For the remainder of the paper we will use both the Pearson ($r_p$) and
Spearman ($r_s$) coefficients as metrics of model accuracy.

\subsubsection{Model sensitivity to the database size}
\label{sec:varysize}

\begin{figure*}[t]
\centering
\subfigure[Sensitivity to the database size.]{
\includegraphics{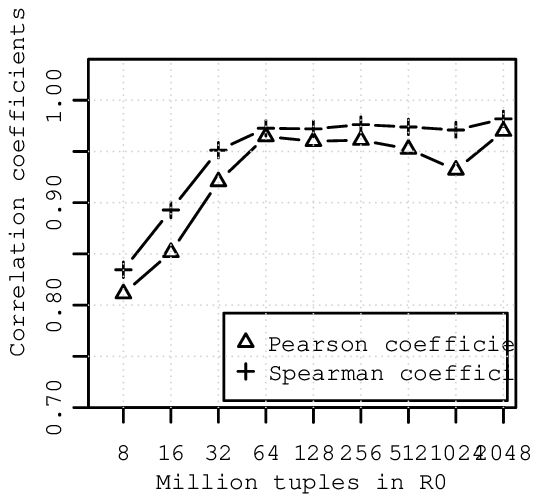}
\label{fig:varysize}
}
\subfigure[Sensitivity to data skew.]{
\includegraphics{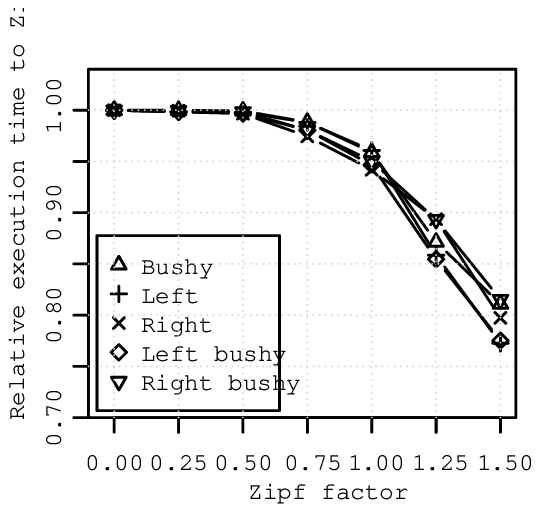}
\label{fig:skew}
}
\subfigure[Sensitivity to join selectivity.]{
\includegraphics{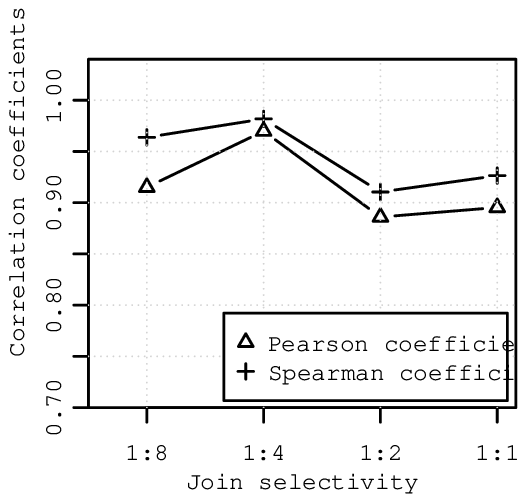}
\label{fig:selectivity}
}
\caption{Model sensitivity to changes in the database 
size, skew, and join selectivity 
for the Intel system.}
\label{fig:modelsensitivity}
\end{figure*}

The proposed model assumes that caching 
affects all query plans equally. 
In this section we validate this design assumption by showing the
accuracy of the model as a growing fraction of the database fits in
the cache (that is, the database shrinks in size).
We use the multi-join query in Section~\ref{sec:workload} on the Intel
system, and fix the join ratio to be 1:4. 
Figure~\ref{fig:varysize} plots the Pearson and Spearman correlation
coefficients as $|R_0|$ ranges from 2
billion tuples to as little as 8 million tuples.
The data points for 2 billion tuples reflect the coefficients for
Figure~\ref{fig:modelour}.
The dip for 1 billion tuples is caused by three outliers, namely the 
``LB2103'', ``L1203'' and ``L2103'' plans, whose response time is in the
slowest quartile (25\%) among all query plans.  These very expensive
query plans would realistically be never considered as viable
alternatives by the query optimizer.
The model remains very accurate with a Spearman coefficient of at least
0.95, until the database size becomes less than 1 GB (that is, when
$R_0$ has fewer than 32 million tuples).
When $|R_0|$ is 8 million tuples, $R_1$, $R_2$, and $R_3$ fit
entirely in cache.
Even then, both the Pearson and Spearman correlation coefficient for our
model remain above 0.8 and 
26 out of the 40 query plans have response times that are within $\pm 15\%$
of the model prediction.

\subsubsection{Model sensitivity to data skew}
\label{sec:varyskew}

Data skew
has been shown to have a profound effect on performance even for a
single join \cite{cagri_hashjoin}.
We now evaluate the effect of data skew on our model's prediction. 
We use the multi-join query discussed in Section~\ref{sec:workload} and
we fix the cardinality of $R_0$, $R_1$, $R_2$ and $R_3$ to be 2 billion,
512 million, 128 million and 32 million tuples, respectively.
Attribute $R_k.b$ ranges from 1 to $|R_{k+1}|$, and the
probability for $R_k.b$ to reference individual keys in $R_{k+1}.a$
follows the Zipf distribution. 
We vary the Zipf factor from 0 (no skew) to 1.5. 

Figure~\ref{fig:skew} plots the normalized query response time of 
query plans that exhibit probe-side data skew.
Since our model doesn't account for data skew, it produces
the same prediction for all Zipf factors. 
The query plan response time remains stable
when the Zipf factor is smaller than 0.5.
When Zipf factor is 1, the most frequent value
for $R_0$ occurs more than 100 million times but the drop in query
response time is less than $6\%$. More skew affects all query plans
almost proportionally for Zipf factors as high as 1.5. 
Given that different query plans have up to $4\times$ different
response times, the relative order between the query plans 
does not change: our model perfectly ranks all the queries
that are shown in Figure~\ref{fig:skew} (that is, $r_s=1.0$).

\subsubsection{Model sensitivity to input cardinality ratio}
\label{sec:varyratio}

\begin{table}[t]
\newcommand{\mc}[3]{\multicolumn{#1}{#2}{#3}}
\centering
\begin{tabular}{ccccc}
\toprule
            & \mc{4}{c}{Cardinality ratio}      \\
\cmidrule(rl){2-5}
Coefficient & 1:1       & 1:2   & 1:4   & 1:8   \\
\midrule
Pearson     & undefined & 0.981 & 0.970 & 0.883 \\ 
Spearman    & undefined & 0.984 & 0.982 & 0.957 \\
\end{tabular}
\caption{Model sensitivity to cardinality ratio.}
\label{tab:varyratio}
\end{table}

We now evaluate the sensitivity of the model as the ratio of 
$\frac{|R_{k+1}|}{|R_{k}|}$ varies from 1 to $\frac{1}{8}$.
When the ratio is 1:1, $|R_0|=|R_1|=|R_2|=|R_3|$ and
$R_k.b$ and $R_{k+1}.a$ have a 1-to-1 match. 
When the ratio is 1:8, $|R_k|=8\times|R_{k+1}|$ and $R_k.b$ and $R_{k+1}.a$
have a 1-to-8 match. 
The correlation coefficients for different ratios are shown in
Table~\ref{tab:varyratio}. 
The coefficient for 1:1 is undefined because our model
gives the same prediction to all 40 query plans and results in a
variance of zero. 
However, the model prediction for the 1:1 ratio is very accurate, as the observed
response time for all 40 query plans varies less than $\pm0.5\%$---well
within the margin of statistical error.
As the ratio changes, the Pearson coefficient stays above 0.88
and the Spearman coefficient stays above 0.95. The drop 
for the 1:8 ratio is caused by three 
outliers (the ``LB2103'', ``L1203'' and ``L2103'' plans) which are among the
most expensive plans for
evaluating this query and thus are not competitive evaluation strategies.

\subsubsection{Model sensitivity to join selectivity}
\label{sec:varyselectivity}

In this section we study the accuracy of the model when changing the
selectivity of the join.
We use the join query in Section~\ref{sec:workload}. The
cardinalities of $R_0$, $R_1$, $R_2$ and $R_3$ are 2 billion, 512
million, 128 million and 32 million respectively.
The results are shown in Figure~\ref{fig:selectivity}.
In the 1:8 dataset, the attribute $R_k.b$ ranges from 1 to
$\frac{|R_{k+1}|}{2}$ and every value occurs 8 times, i.e.~half of the
tuples in $R_{k+1}$ will have matching tuples in $R_k$ and each
$R_{k+1}$ tuple will match 8 tuples in $R_0$. The other datasets are
obtained likewise: for instance, in the 1:1 dataset only $\sfrac{1}{4}$ of the tuples in
$R_{k}$ will match in $R_{k+1}$, and each will match exactly once.
As Figure~\ref{fig:selectivity} shows, the model remains accurate as the
join selectivity changes and the Spearman coefficient stays above 0.9.

\begin{figure}
\centering
\subfigure[Observed performance.]{
\includegraphics{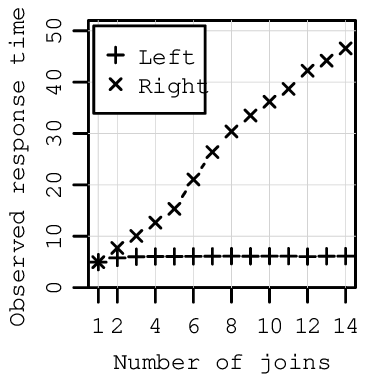}
\label{fig:longpipeline:obs}
}
\subfigure[Predicted performance.]{
\includegraphics{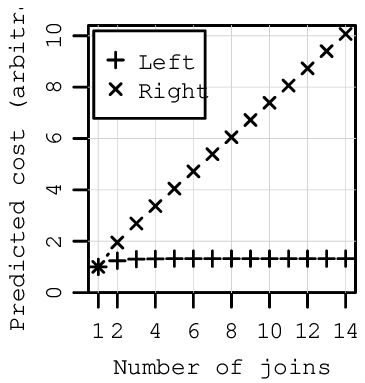}
\label{fig:longpipeline:pred}
}
\caption{Comparing the optimal left-deep and right-deep query
plans as the number of joins increases. 
The performance gap grows to almost $10\times$, as predicted by the
model.
}
\label{fig:longpipeline}
\end{figure}

\subsubsection{Modeling longer join pipelines}
\label{sec:morejoin}
After focusing on the exhaustive exploration of the query plan space for
a query with three joins, we now will explore how the performance of the
best right-deep tree (``R3210'') and the best left-deep tree (``L3210'')
changes as the number of joins increases, and whether our model can
predict this accurately.
We generate more relations with a join ratio of 1:4 
as in Section~\ref{sec:diskmodel}, and we grow each query tree
accordingly.
We plot the query response time when the number of joins increases for
both the left-deep tree and the right-deep tree in
Figure~\ref{fig:longpipeline:obs}, and contrast this with the 
result from our prediction in Figure~\ref{fig:longpipeline:pred}. 
The left-deep tree and the right-deep tree have the same performance
when the number of joins is 1, as the trees are identical. As the
number of joins increases, the performance of right-deep
trees linearly worsens over the performance of the left-deep tree. 
The key drawback of the right-deep
tree is that by using the largest relation as the probing relation
the data of the largest relation ``flows'' through all pipeline stages
and results in additional \rr memory accesses in every stage.
Our model accurately forecasts this performance trend for left-deep
trees and right-deep trees as the number of joins increases.

\section{Concluding remarks}
\label{sec:conclusion}

A significant fraction of a
database may fit entirely in main memory and can be analyzed in
parallel using multiple CPU cores to keep query response times short.
Accurately forecasting the response time of different in-memory query plans 
is becoming important for query optimization in this environment. 
Towards this goal, we contribute a cost model that can accurately
predict the query response time of ad-hoc query plans with multiple
hash-based joins.

A surprising insight from our model is that some left-deep
query trees can be more efficient than their bushy and right-deep tree
counterparts because they result in less memory I/O during execution.
Prior work in parallel hash-based multi-join query evaluation has
advocated right-deep query trees because all build subtrees can be
processed in parallel and then probed in a single pipeline
\cite{dono:tradeoff}.
Our experimental evaluation corroborates that in a main-memory setting
evaluating query subtrees sequentially using all threads is as fast as
evaluating separate query subtrees concurrently. 
This finding suggests that evaluating one join at a time and storing the
intermediate result in a hash table may be a
viable query execution strategy over a memory-resident dataset as it can
ameliorate the cascading effect of errors in join cardinality
estimation. 
Query processing techniques that are adaptive to cardinality estimation
errors are a promising research area for future work.

\section*{Acknowledgements}
We would like to acknowledge Tasos Sidiropoulos,
Yang Wang, the anonymous reviewers and our shepherd, Jennie Duggan,
for their actionable and insightful comments that improved this paper.
This research was partially supported by the National Science Foundation
under grant III-1422977 and by a Google Research Faculty Award.
The evaluation was conducted in part at the National Energy Research
Scientific Computing Center, a DOE Office of Science User Facility
supported by the Office of Science of the U.S. Department of Energy
under Contract No. DE-AC02-05CH11231.

\bibliographystyle{abbrvnat}
\bibliography{refer}

\end{document}